\documentclass[fleqn,twoside]{article}
\usepackage[headings]{espcrc2}

\readRCS
$Id: espcrc2.tex,v 1.2 2004/02/24 11:22:11 spepping Exp $
\usepackage{graphicx}
\usepackage[figuresright]{rotating}

\newcommand{\ttbs}{\char'134}
\newcommand{\AmS}{{\protect\the\textfont2
  A\kern-.1667em\lower.5ex\hbox{M}\kern-.125emS}}

\vspace{-3.0cm}
\title{Exclusive production of vector mesons in $\gamma p$ and $p p$ collisions.}

\author{A. Cisek\address[IFJ]{Institute of Nuclear Physics PAN, PL-31-342 Cracow, Poland
},
        W. Sch\"afer\addressmark[IFJ],
        \thanks{For following authors with the same
                address use the {\tt\ttbs addressmark} command.},
        A. Szczurek\addressmark[IFJ] $^{,}$
                \address{University of Rzesz\'ow, PL-35-959 Rzesz\'ow,Poland}
}

\runtitle{Exclusive production of vector mesons in $\gamma p$ and $p p$ collisions.}
\runauthor{A.~Cisek {\it et al.}}

\begin{document}

\footnotesize{
\begin{abstract}
The dominant mechanism for the central exclusive production of vector
meson in $p p$ and $p \bar p$ collisions is the \\
$\gamma$ $I\!\! P$ fusion. As a building block for the pp reaction, the
amplitude for photoproduction $ \gamma p \to V p $ is calculated
in a pQCD $k_{T}$- factorization approach. We will present results for
several vector mesons: $\rho$, $\omega$, $\phi$, $J/\Psi$ and $\Upsilon$.
The total cross section for diffractive mesons production as a function
of energy and photon virtuality is calculated. We will present dependence
on the mass of the quark for light mesons.
The results for $ \gamma p \to V p $ photoproduction depend on the model
of the meson wave function.
We compare our results with  a HERA data for photon-proton collisions.
Finally we turn to pp collisions, and we present distribution in rapidity,
transverse momentum of vector mesons and azimuthal angle between outgoing protons
for Tevatron and LHC energies. The absorption effects will be discussed.
\vspace{1pc}
\end{abstract}
}

\maketitle

\vspace{-3.5cm}
\footnotesize{
\section{Introduction}
Photoproduction of the vector mesons in photon-proton collisions is
interesting from both experimental and theoretical side. It was studied intensively
by many people. Photoproduction process $\gamma p \to V p$ has been masured
at HERA. When calculating the cross section for photoproduction
at high energies, the two main ingredients are
the unintegrated gluon distribution function and the quark-antiquark
wave function of the vector meson. Photoproduction of vector mesons can be also studied
in proton-proton or proton-antiproton collisions, where it is the dominant
mechanism of exclusive production of vector mesons at central rapidities.
We refer to this production mechanism also as photon-Pomeron fusion \cite{RSS}.
For an evaluation of differential distributions
it is important to include the effect of absorptive corrections.
The HERA data on photoproduction of vector mesons constrain the exclusive
production at Tevatron for not too large rapidity of the vector meson.
}

\vspace{-0.2cm}
\section{Photoproduction $\gamma p \longrightarrow Vp$ at HERA}

\begin{figure}
\begin{center}
  \includegraphics[height=0.15\textheight]{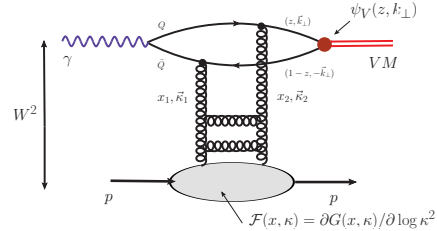}
\vspace{-1.0cm}
  \caption{\small{A sketch of the amplitude for exclusive photoproduction
            $\gamma p \to V p$.}}
\label{VM_amplitude}
\end{center}
\end{figure}

The amplitude for the reaction is shown schematically in Fig.\ref{VM_amplitude}.
The full amplitude for this process can be writen as (see Refs.\cite{RSS,INS06,IN02}):
\footnotesize{
\begin{eqnarray}
{\cal M}_{L,T}(W,\Delta^2,Q^2) = (i + \rho_{L,T}) \, \Im m {\cal M}_{L,T}(W,\Delta^2=0,Q^2)\times
\nonumber \\
\exp{\frac{(-B(W) \Delta^2)} {2}} \, ,
\end{eqnarray}}
\small{where $\rho_{L,T}$ is a ratio of real and imaginary part of the
amplitude for longitudinal and transverse polarization of the photon and
$B(W)$ is slope parameter which depends on energy:
\small{$B(W) = B_0 + 2 \alpha'_{eff} \log \Big( {\frac {W^2} {W^2_0}} \Big)$}.
\small{
We have different $B_0$ for different mesons. This values can be found in Refs.
\cite{RSS,CSS_Phi,CLSS,SS07,CSS}.
The imaginary part of the amplitude depends on the unintegrated gluon distibution
function and on the wave function of the vector meson. The explicit form of the amplitude
for longitudinal and transverse polarrization can be found in Refs. \cite{INS06,RSS}.

Our amplitude is normalized to the cross section:}
\small{
\begin{eqnarray}
\sigma_{L,T}(\gamma p \to V p) = \frac{1 + \rho_{L,T}^2} {16 \pi B(W)} \, \Big| {\frac {\Im m { {\cal M}_{L,T}(W,\Delta^2)}} {W^2} } \Big|^2 \, .
\end{eqnarray}

\small{
We calculated separetly cross section for transverse ($\sigma_{T}$)
and longitudinal ($\sigma_{L}$) polarizations. The full cross section
is a sum of these two components. In our calculation we used two
types of model wave functions, Gausian:}
\small{
\begin{eqnarray}
\psi_{1S}(p^2) = C_1 
\exp \Big( - \frac{p^2 a_1^2}{2} \Big) \, ,
\nonumber
\end{eqnarray}
\begin{eqnarray}
\psi_{2S}(p^2) = C_2 \Big( \xi_0 - p^2 a_2^2 \Big) 
\exp \Big( - \frac{p^2 a_2^2}{2} \Big)
\end{eqnarray}}
\normalsize{and Coulomb--like wave functions:}
\small{
\begin{eqnarray}
\psi_{1S}(p^2) = {C_1 \over \sqrt{M}} \, {1 \over (1 + a_1^2 p^2)^2} \,  ,
\nonumber \\
\psi_{2S}(p^2) = {C_2 \over \sqrt{M}} \, {\xi_0 - a_2^2 p^2 \over (1 + a_2^2 p^2)^3} \, .
\end{eqnarray}
}
\small{
The parameters of the wave function are obtained from fitting the decay widths into
$e^{+}$ $e^{-}$.}

\vspace{-0.2cm}
\subsection{Numerical results and HERA data}

\begin{figure}[!htp]
\begin{center}
\includegraphics[width=0.3\textwidth]{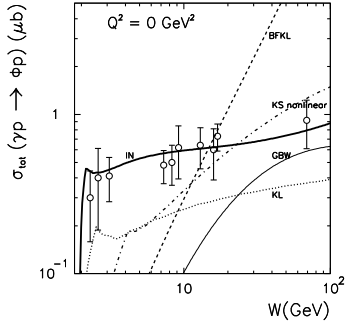}
\vspace{-1.0cm}
\caption{\small{Total cross section for $\gamma p \longrightarrow \phi p$
as a function of energy. Results for different models UDDF function.
Experimental data can be found in Refs.\cite{ZEUS96,Ferm_Phi}.}}
\label{Phi_tot}
\end{center}
\end{figure}

\small{
In Figs.\ref {Phi_tot} the total cross section for photoproduction
$\gamma p \longrightarrow \phi p$ is shown as a function of photon-proton
center of-mass energy for $Q^{2}=0$. We present results for differnt models of UGDF function.
The thick solid line is for the Ivanov-Nikolaev model, the dash-dotted line is for
the Kutak-Sta\'sto model, the dased line is for the BFKL, the dotted line is for
the Kharzeev-Levin UGDF and the thin solid line is for the Golec-Biernat-W\"usthoff model.
We can see that Ivanov-Nikolaev UGDF the best describes experimental data.
}

\begin{figure}[!h]
\begin{center}
\includegraphics[width=0.3\textwidth]{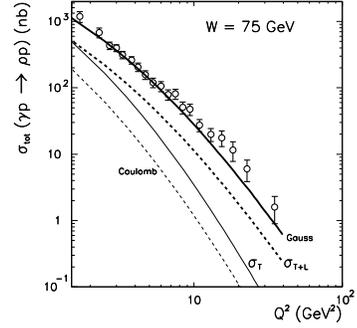}
\vspace{-1.0cm}
   \caption{\label{Q2_rho}
   \small{Cross section as a function of $Q^2$ for photoproduction ($W =$ 75 GeV)
    mesons rho. Experimental data are taken from Ref.\cite{H1_rho}.}}
\end{center}
\end{figure}

\small{
The total cross section at finite $Q^2$ is a sum of transverse and longitudinal
components.
In Fig.\ref{Q2_rho} we present total and transverse cross section for the energy
$W =$ 75 GeV. This cross section is a function of $Q^2$. The thick lines are for
the Gaussian wave function and thin lines are for Coulomb wave function.
The solid lines are for total cross section and the dashed lines are for transverse
cross section. We compare our results with experimental data of the ZEUS Collaboration.
}

\begin{figure}[!htp]
\begin{center}
\includegraphics[width=0.3\textwidth]{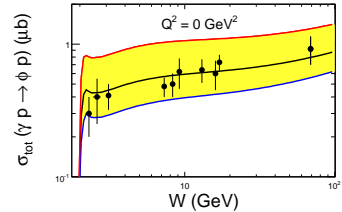}
\vspace{-1.0cm}
\caption{\small{Total cross section for $\gamma p \longrightarrow \phi p$
as a function of energy. Different values of the strange quark mass.
Experimental data can be found in Refs.\cite{ZEUS96,Ferm_Phi}.}}
\label{Phi_mass}
\end{center}
\end{figure}

\small{
In Fig.\ref {Phi_mass} the total cross section as a function of photon-proton
center of-mass energy for $Q^{2}=0$ . We show results for three different values of
the strange quark mass. The red (upper) line is for $m_s =$ 0.37 GeV,
blue (lower) line for $m_s =$ 0.50 GeV and the black line (which goes
throught the data points) for $m_s =$ 0.45 GeV. We can see that the results
for $m_s =$ 0.45 GeV give the best description of the experimental data..
}

\begin{figure}[!h]
\begin{center}
\includegraphics[width=0.3\textwidth]{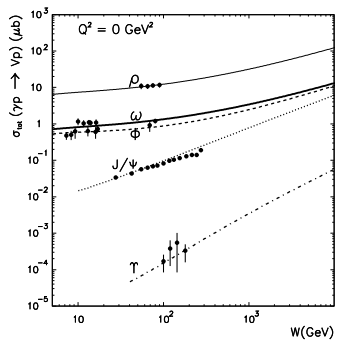}
\vspace{-1.0cm}
   \caption{\label{w_sig_VM}
   \small{Total cross section as a function of energy for photoproduction
          ($Q^{2} = 0\ GeV^{2}$) process ($\gamma p \to V p$).
          Experimental data are taken from Refs.\cite{ZEUS96,Ferm_Phi,H1_rho,ZEUS_Ups}.}}
\end{center}
\end{figure}

\small{
In Fig.\ref{w_sig_VM} we present total cross section for different mesons. This cross section
is a function of energy. We can see that for higher energy ($>$ 5 TeV) total cross sections for
$J/\Psi$, $\phi$ and $\omega$ are very similar. The cross section for $\rho$ meson is
much bigger in the energy range considered here. Here we have presented results for the 
Gaussian wave function.
}

\vspace{-0.2cm}
\section{Exclusive photoproduction in $p \bar p $ collisions}

\begin{figure}
  \includegraphics[height=.12\textheight]{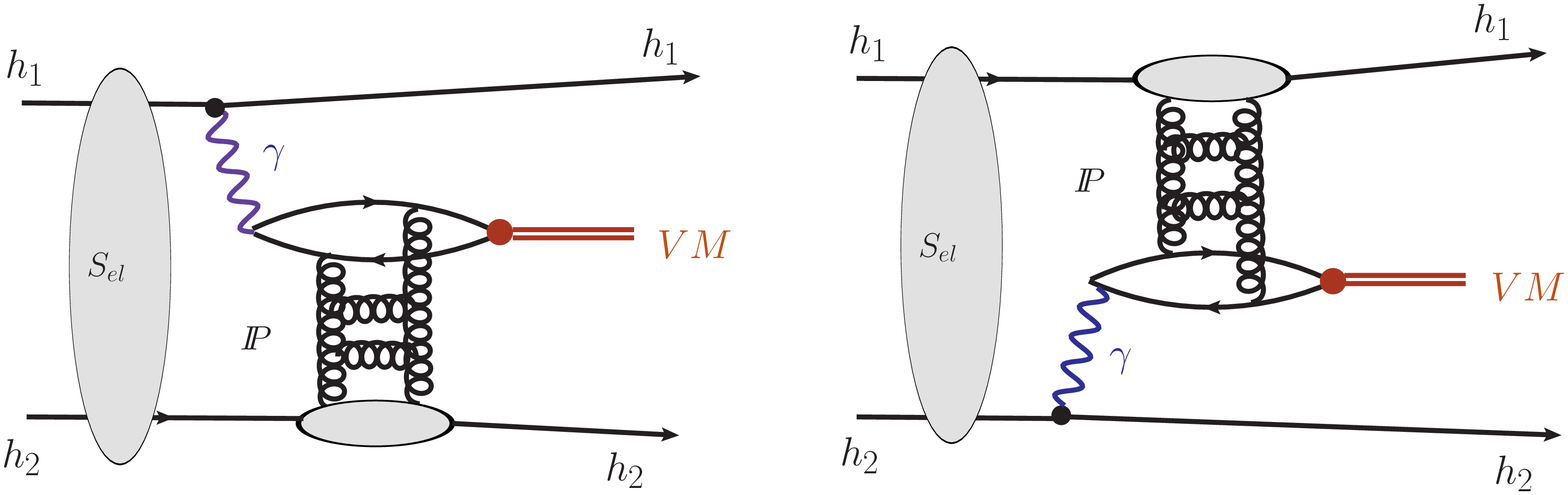}
\vspace{-1.0cm}
  \caption{\small{A sketch of the exclusive  $p p \to p V p$.}}
\label{amplitude_abs}
\end{figure}

\small{
The diagrams in (Fig.\ref{amplitude_abs})
show schematically the amplitude with absorptive correction, including elastic
rescattering. The full amplitude for $ p p \to p V p$ or $ p \bar p \to p V \bar p$
can be written as \cite{RSS,SS07}:
}

\small{
\begin{eqnarray}
M(p_1,p_2) = \int{\frac{d^2 k} {(2 \pi)^2}} S_{el}(k) M^{(0)}(p_1 - k, p_2 + k) =
\nonumber \\
M^{(0)}(p_1,p_2) - \delta M(p_1,p_2)\, .
\label{rescattering_term}
\end{eqnarray}
}

\small{
In formula (\ref{rescattering_term}) $M^{(0)}(p_1,p_2)$ is the Born-amplitude
(without absorption) for the process $ p p \to p V p$ or $ p \bar p \to p V \bar p $
which includes our amplitude for photoproduction and $\delta M(p_1,p_2)$
is the absorptive correction. We have calculated our amplitude for the Ivanov-Nikolaev unintegrated gluon
distribution function and the Gaussian wave function. In formula (\ref{rescattering_term})
$p_1$ and $p_2$ are transverse momenta of outgoing protons.
The differential cross section is given in terms $M$ as:
}
\small{
\begin{eqnarray}
d \sigma = {\frac{1} {512 \pi^4 s^2 }} | M |^2 \, dy dt_1 dt_2 d\varphi
\, ,
\end{eqnarray}}
\small{where $\varphi$ is azimuthal angle between outgoing $p p$ or $p \bar p$.
}

\vspace{-0.2cm}
\subsection{Numerical results for proton-proton and proton-antiproton collisions}

\begin{figure}[!htp]
\begin{center}
\includegraphics[width=3.1cm]{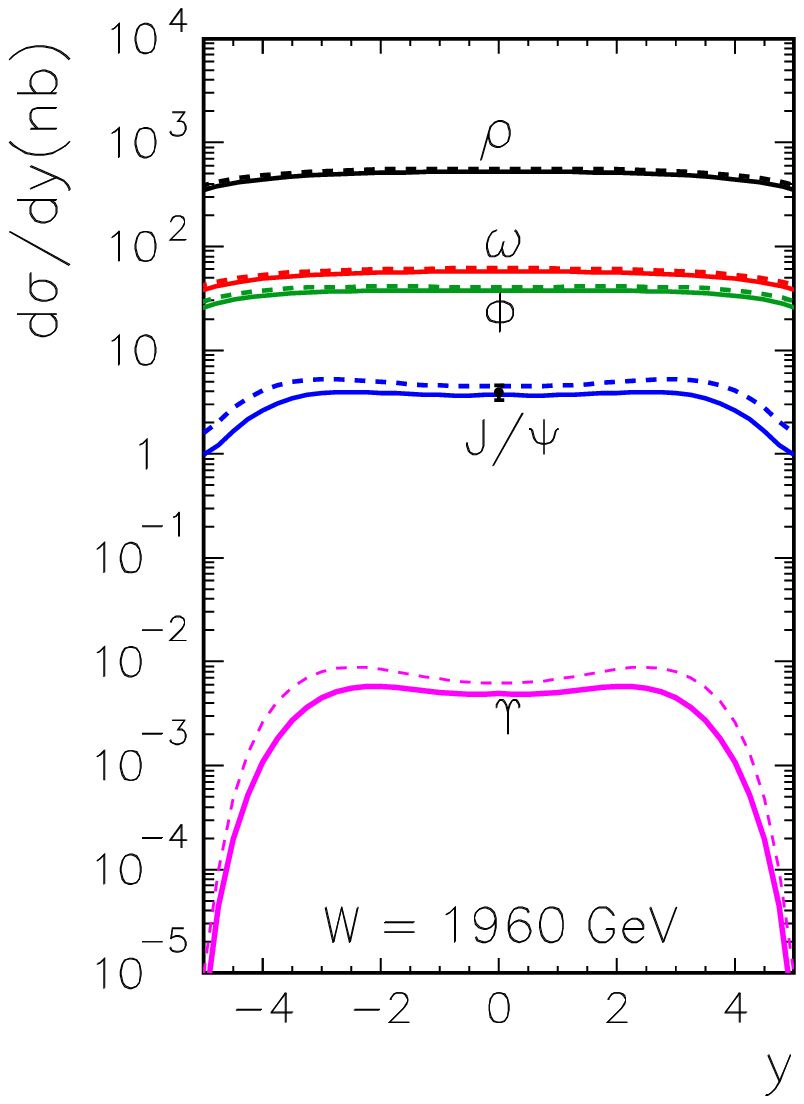}
\includegraphics[width=3.1cm]{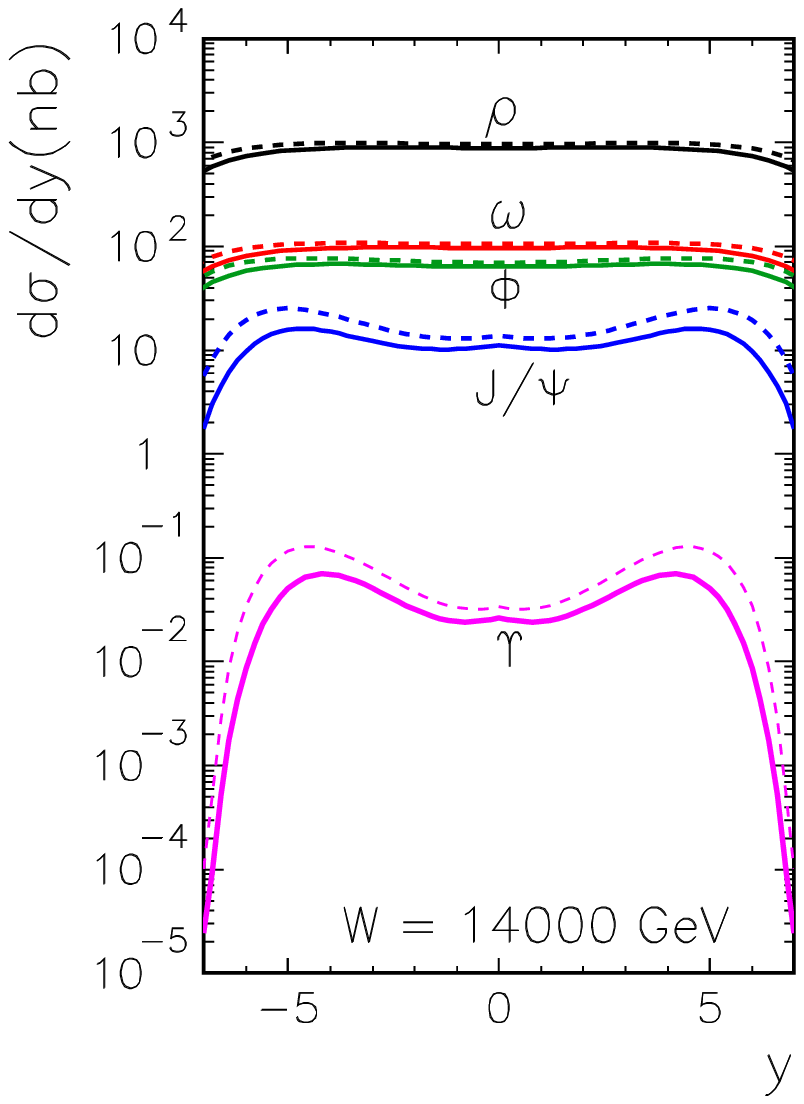}
\vspace{-1.0cm}
\caption{\small{Rapidity spectrum of vector mesons for Tevatron
(left panel) and LHC (right panel) energy {\protect \cite{CDF}}.}}
\label{rapidity}
\end{center}
\end{figure}

\small{
In Fig.\ref {rapidity} we show rapidity distribution for various vector meson in proton-antiproton collisions.
Our results are compared with recent CDF data \cite{CDF} for $J/\Psi$. The solid lines are
for the amplitude with absorptive corrections and the dashed lines are for the amplitude without absorption.
}

\begin{figure}[!ht]
\begin{center}
\includegraphics[width=0.23\textwidth]{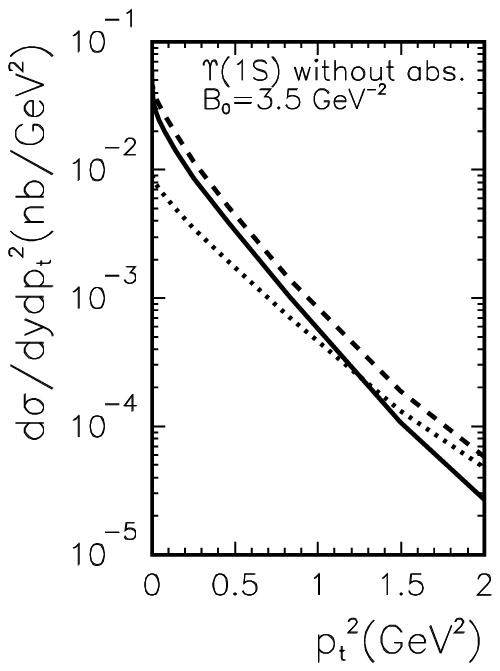}
\includegraphics[width=0.23\textwidth]{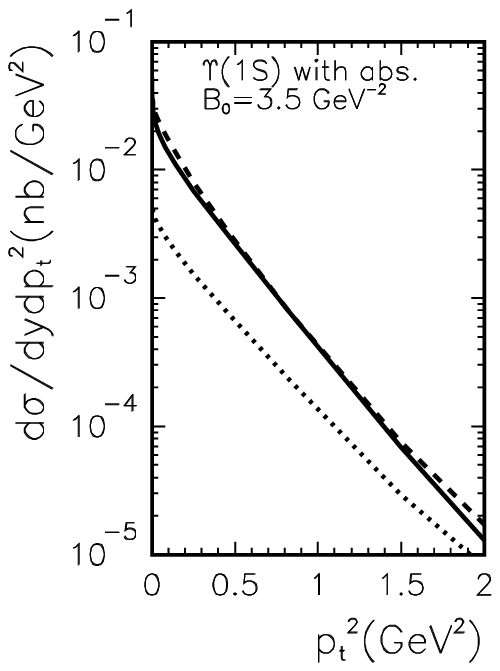}
\includegraphics[width=0.23\textwidth]{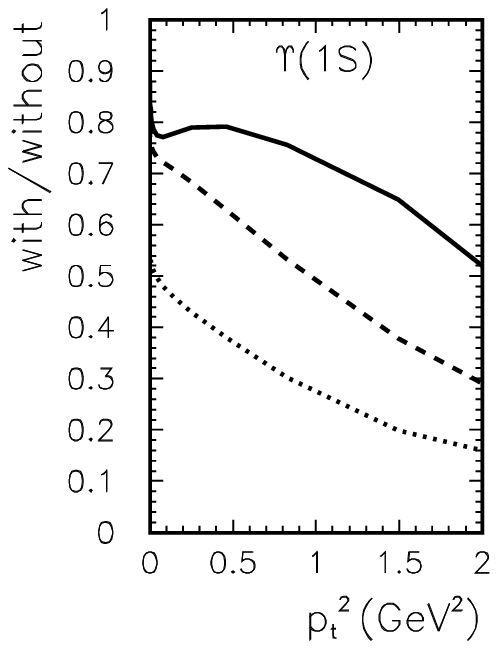}
\vspace{-1.0cm}
\caption[*]{\small {Invariant cross section $d\sigma/dydp_t^2$ as a function
of $p_t^2$ for $\Upsilon$(1S) at Tevatron energy. Left (upper): without absorption;
Right (upper): with absorptive corrections. Center (lower): Ratio of cross sections
with absorptive corrections included/switched off.}}
\label{pt}
\end{center}
\end{figure}

\small{
In Fig.\ref{pt} we show distributions in transverse momentum for $\Upsilon$
at the Tevatron energy: $y=0$ (solid), $y=2$ (dashed) and $y=4$ (dotted) for
different values of rapidity. We present results for bare amplitudes (left - upper)
and for the amplitudes with absorptive corrections (right - upper).
We show the ratio of the invariant cross section with to without absorptive
corrections (center - lower). We can see that absorption effects depend on rapidity and $p_{t}$.
}

\begin{figure}[!h]
\includegraphics[width=0.23\textwidth]{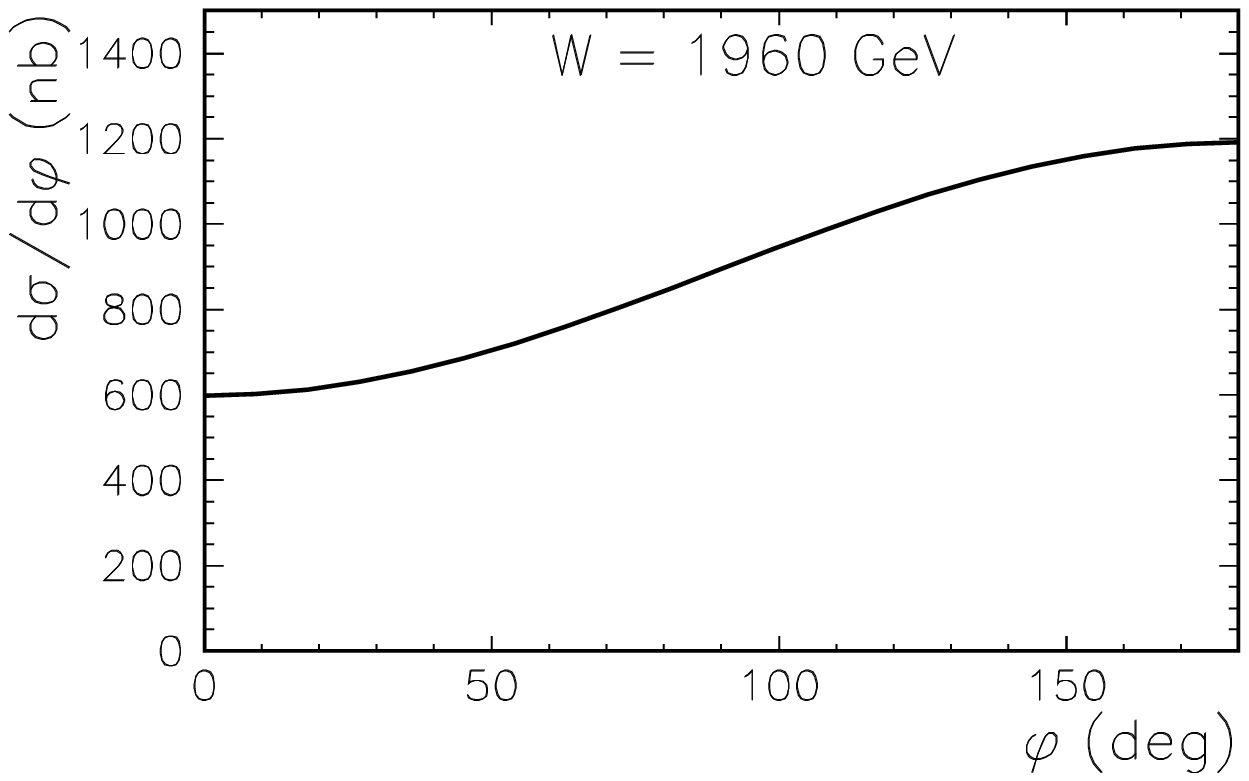}
\includegraphics[width=0.23\textwidth]{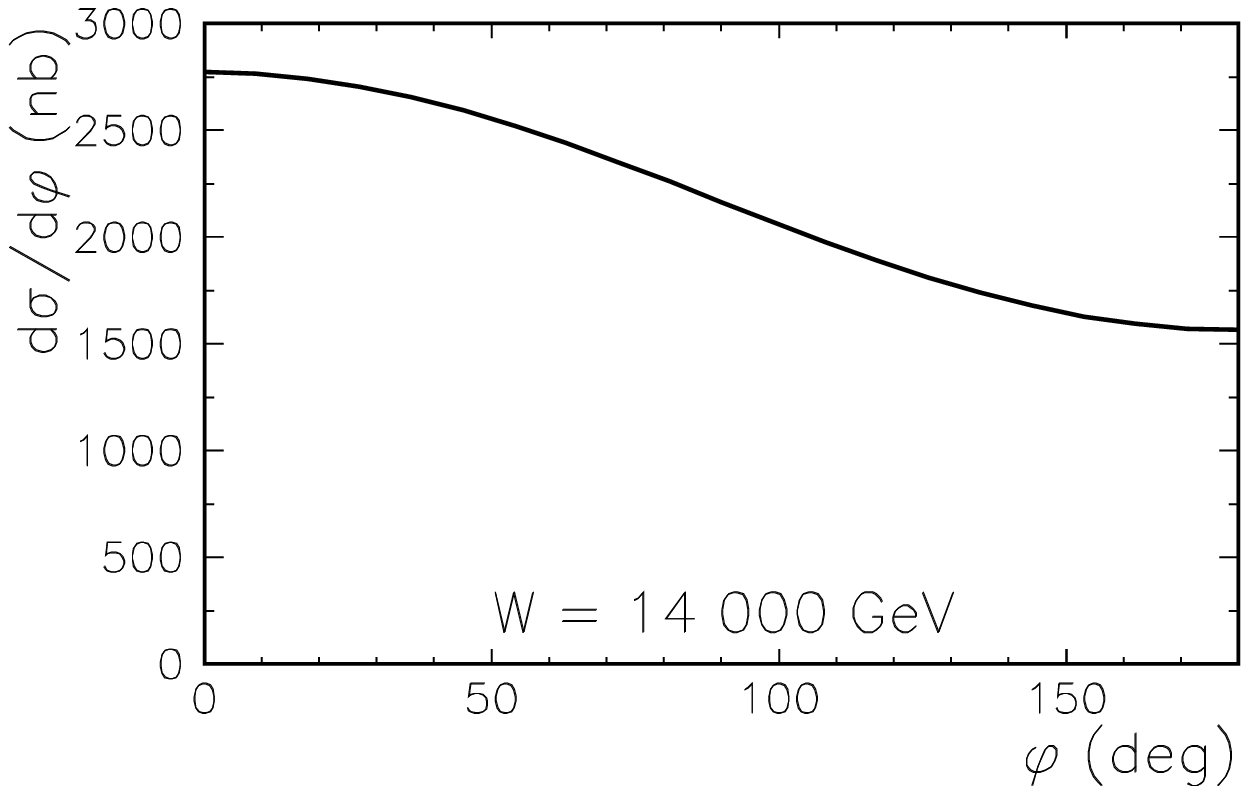}
\vspace{-1.0cm}
  \caption{\label{dsig_dphi}
   \small Distribution in azimuthal angle for $\rho$ meson production at the Tevatron (left panel)
          and LHC (right panel) energies.}
\end{figure}

\small{
In Fig.\ref{dsig_dphi} we show the distribution in relative azimuthal angle between
outgoing protons. The distributions are for the LHC ($pp$ collisions)
and Tevatron ($p\bar p$ collisions). The dependence on $\varphi$ comes from the interference
between $\gamma I\!P$ and $I\!P \gamma$ components. The interference is different for LHC ($pp$)
and Tevatron ($p\bar p$) because proton and antiproton have opposite charges.
}

\vspace{-0.2cm}
\section{Conclusions}

\footnotesize{
We have calculated the total cross section for diffractive vector meson photoproduction
$\gamma p \to V p$ in a pQCD-based model for $\rho$, $\omega$, $\phi$, $J/\Psi$
and $\Upsilon$. The results for photoproduction $\gamma p \to V p$ depend on the model
of the wave function and UGDFs function. The Gauss wave function better describes data than the Coulomb
one. We can see that the Ivanov-Nikolaev unintegrated distribution the best describes experimental data.
We have compared our results with a recent HERA data.
Based on these photoproduction amplitudes, we have predicted cross sections for exclusive
production of vector mesons in $p p$ and $p \bar p$ collisions.
In our calculation of hadronic processes we have included explicitly absorption effects.
This effect depends on rapidity and $p_t$.
}





\end{document}